\pdfoutput=1
\documentclass[aps,prd,twocolumn,superscriptaddress,nofootinbib,floatfix,amsmath,amssymb,longbibliography]{revtex4-2}

\usepackage{graphicx}
\usepackage{xcolor}
\usepackage{hyperref}

\makeatletter
\AtBeginDocument{%
  \def\label#1{\@bsphack
    \protected@write\@auxout{}{%
      \string\newlabel{#1}{{\@currentlabel}{\thepage}{}{\@currentHref}{}}%
    }%
    \@esphack}%
  \let\ltx@label\label
}
\makeatother

\providecommand{\jcap}{J. Cosmology Astropart. Phys.}
\providecommand{\apj}{ApJ}

\providecommand{\aap}{A\&A}

\providecommand{\mnras}{MNRAS}
\providecommand{\prd}{Phys. Rev. D}
\providecommand{\prl}{Phys. Rev. Lett.}

\providecommand{\aj}{AJ}

\defcitealias{DesmondFerreira2020}{DF20}
\newcommand{\DF}{\citetalias{DesmondFerreira2020}}

\begin{document}

\title{Galactic warps constrain equivalence-principle violation in the dark sector}

\author{Harry Desmond}
\email{harry.desmond@port.ac.uk}
\affiliation{Institute of Cosmology \& Gravitation, University of Portsmouth,
Dennis Sciama Building, Portsmouth PO1 3FX, UK}

\author{Pedro G. Ferreira}
\affiliation{Astrophysics, University of Oxford, Denys Wilkinson Building,
Keble Road, Oxford OX1 3RH, UK}

\begin{abstract}

Current measurements of the expansion of the Universe allow for the presence of evolving dark energy, distinct from a cosmological constant, with an equation of state that seems to violate the null energy condition. One notable explanation for this behaviour is that the dark energy interacts with the dark matter, violating the equivalence principle yet evading conventional laboratory and astrophysical constraints. Such models produce fifth forces acting only on dark matter. We show how such forces separate halo and stellar disk centres, creating a potential gradient across the disk that warps it into a cup shape.
Measuring the warps of $4\times10^3$ edge-on disks in Sloan Digital Sky Survey images and modelling the fifth-force signal together with competing baryonic effects, we find no sign of the predicted correlation between the measured warps and the fifth-force field. This constrains the fifth-force strength relative to gravity to $\Delta G/G_N<7.7\times10^{-4}$ at 95\% credibility (or $<1.7\times10^{-3}$ once systematic uncertainties are included) for force ranges of $8$\,Mpc and above.
In the infinite-range limit of a scalar rolling today, even the conservative bound is three times stronger than the tightest cosmological bound (which, unlike ours, depends on the scalar's cosmological evolution) and about two orders of magnitude below other astrophysical bounds. It disfavours at $2$--$3\sigma$ the couplings preferred by recent fits of interacting dark energy to cosmological data unless gravitational screening is invoked.

\end{abstract}

\date{\today}
\maketitle

\section{Introduction}\label{sec:intro}

Recent cosmological observations have suggested that the expansion of the Universe may not be driven by a cosmological constant, $\Lambda$, but by some form of dynamical dark energy \cite{DESI2024VI}. The preference is for the equation of state of this dark energy to be ``phantom''-like, i.e.\ $w<-1$, for at least part of its evolution.

The large majority of proposals for dark energy involve a late-time evolving scalar field \cite{Ratra:1987rm, Peebles:1987ek,Ferreira:1997au,Ferreira:1997hj, Caldwell:1997ii, Perrotta:1999am, Amendola2000, Brax:1999gp, Barreiro:1999zs, Sahni:1999gb}. If the dark energy is to have $w<-1$ and satisfy the equivalence principle, it generically  requires fifth forces  \cite{Wolf2025,Garcia-Garcia:2026nzy}. These are strongly constrained on small scales \cite{adelberger2003tests}, necessitating gravitational screening which complicates the structure and tests of the theories.

Alternatively, the dark energy may couple {\it only} to the dark matter, leading to an  exchange of energy between the two \cite{Wetterich1988,Amendola2000,Copeland2006}. Such an interaction violates the equivalence principle in the dark sector, but allows dark energy to mimic phantom behaviour while the total dark sector still satisfies the null energy condition. This possibility is an old one \cite{Frieman1991,Gradwohl1992}, and has even been argued to be positively desirable: a dark-sector force of strength similar to gravity empties voids and alters baryon fractions in ways that have been claimed to improve on $\Lambda$ cold dark matter ($\Lambda$CDM) \cite{GubserPeebles2004,Keselman2009}.

The equivalence principle is among the most precisely tested propositions in physics, with torsion-balance, lunar-laser-ranging and space tests reaching parts in $10^{13}$ to $10^{15}$ \cite{Schlamminger2008,Williams2004,Touboul2022}, but these tests are insensitive by construction to a force acting only in the dark sector.
Here, the tightest bounds are cosmological and arise mainly from the scalar's effect on the expansion history. For a scalar lighter than $H_0$, the cosmic microwave background and baryon acoustic oscillations give $\Delta G/G_N<5.4\times10^{-3}$ \cite{Archidiacono2022}, or $4.8\times10^{-3}$ with the full shape of the galaxy power spectrum \cite{Bottaro2024}. A heavier scalar, of range tens to hundreds of Mpc, is bounded more tightly, at $\Delta G/G_N\lesssim10^{-3}$, mainly through the energy density of its oscillations \cite{Bottaro2025}. Previous astrophysical tests are weaker still:
$7\%$ of gravity from the dark-matter Euler equation \cite{Bonvin2025}, $\sim\!10\%$ from
the Sagittarius stream \cite{Kesden2006}, $\sim\!60\%$ from Fornax \cite{Zamlung2026} and
$7$--$14\%$ from cluster redshifts \cite{DiDio2026}; see also \cite{BovyFarrar2009,Bahrami2019}.
This has permitted readings of the latest data as \emph{favouring} a dark-sector force at a few parts in $10^3$ of gravity, which could ameliorate tensions between different probes: from Dark Energy Spectroscopic Instrument (DESI) baryon acoustic oscillations \cite{Bottaro2025}, from their combination with the galaxy power spectrum \cite{Clini2026}, and, in the swampland framing discussed in Sec.~\ref{sec:implications}, from DESI DR2 with supernovae \cite{Bedroya2025}. 

\begin{figure}[tb]
  \centering
  \includegraphics[width=\columnwidth]{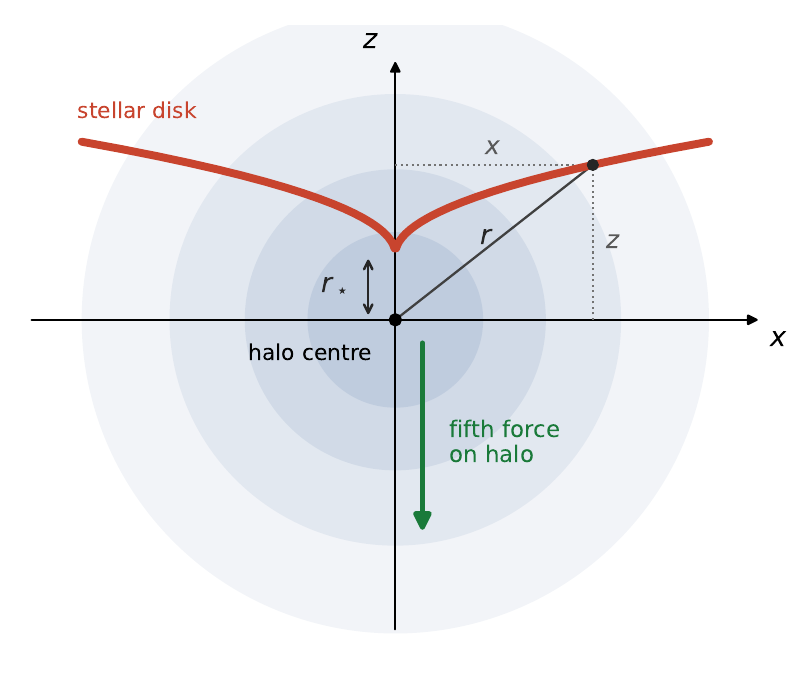}
  \caption{Cartoon illustration of the geometry.
  A fifth force
  acting on dark matter alone accelerates the halo (downwards in this example), causing an offset $r_\star$  between the disk and halo centres, hence bending the disk into a warp.
}
  \label{fig:schematic}
\end{figure}

Dark-sector-only fifth forces \emph{do} however manifest in galaxies --- where baryons set the light and dark matter the dynamics --- so that a dark-sector violation of the equivalence principle becomes visible in galaxy structure \cite{Kesden2006,Kesden2006b,Keselman2009}. The signature we use here, a U-shaped warp of the stellar disk, was identified by \cite{JainVanderPlas2011} for a screened fifth force acting on dark matter and gas but not stars. The geometry is the same for a force on dark matter alone.
As illustrated in Fig.~\ref{fig:schematic}, a force coupling to dark matter but not stars accelerates a galaxy's halo towards surrounding structure and leaves
the disk behind. The disk is bound to the halo, so it settles at the offset where the halo's own
gravity supplies the acceleration the stars are missing: a few tens of parsecs in a galaxy tens
of kiloparsecs across. The restoring pull weakens outwards, so
equilibrium requires a larger displacement at larger radius and the disk bows away from the force.
That bend accumulates across the disk and is in principle measurable in ordinary imaging, making warp amplitude
a fifth-force detector exactly proportional to the force. Since the acceleration is sourced
by all the dark matter within a few times the force range, tens of megaparsecs, a reconstruction of
the local density field predicts each galaxy's warp in size and direction. While the predicted warp is too small to be seen directly per galaxy --- and is swamped by confounding baryonic effects --- a statistical sample of thousands of galaxies combined with a method for marginalising over the baryons can lead to stringent constraints
\cite{DesmondWarps2018,DesmondOffsets2018,DesmondFerreira2020}.

A comprehensive analysis of this effect was presented in \cite{DesmondFerreira2020} (hereafter \DF), which selected Sloan Digital Sky Survey (SDSS) galaxies, measured their warps and constructed a Bayesian forward model to infer the strength and range of a fifth force while marginalising over observational and astrophysical noise. The treatment of baryonic warps as noise was subsequently validated against the Horizon-AGN simulation in \cite{Bartlett2021}, which showed that the warps produced by baryonic effects in that simulation are not significantly degenerate with the fifth-force prediction. \DF\ targeted a thin-shell-screened fifth force, such as arises in $f(R)$ gravity, that couples to gas and dark matter but not stars; although it noted that the same warp test applies to an unscreened fifth force acting only in the dark sector, it presented that case only as a foil.
We take that case up here, reusing the galaxy sample and fifth-force forward model of \DF\ but measuring the warps afresh from the survey images with a more robust algorithm, extending the model to include the disk's self-gravity and the effects of edge-on projection and atmospheric seeing, and adopting a likelihood that accommodates outliers. These refinements somewhat weaken, but make much more robust, the bound that \DF\ quotes based on warps for this case.

The structure of the paper is as follows. Sec.~\ref{sec:coupling} introduces interacting dark energy and the fifth force it produces, Sec.~\ref{sec:data} the galaxy sample and the warp measurement, and Sec.~\ref{sec:model} the forward model and likelihood. Sec.~\ref{sec:bound} presents the bound, Sec.~\ref{sec:screening} discusses screening and Sec.~\ref{sec:caveats} the remaining systematic uncertainties. Sec.~\ref{sec:implications} compares our bound with the literature and Sec.~\ref{sec:conclusion} concludes.

\section{Interacting Dark Energy}\label{sec:coupling}
Let $T_{\rm DM}^{\mu\nu}$ and $T_{\rm \varphi}^{\mu\nu}$ denote the stress-energy tensors of dark matter and dark energy respectively and assume that they are jointly covariantly conserved, $\nabla_\nu(T_{\rm DM}+T_{\varphi})^{\mu\nu}=0$ (where $\nabla_\mu$ is the covariant derivative associated with a metric $g_{\mu\nu}$). One can have energy transfer between the two  \cite{Amendola2000} if one assumes
\begin{equation}
\nabla_\nu T_{\rm DM}^{\mu\nu} = Qu^\mu , \qquad \nabla_\nu T_{\varphi}^{\mu\nu}  = -Qu^\mu ,
\label{eq:generalcoupling}
\end{equation}
where $Q$ is a scalar function and $u^\mu$ is a timelike vector. 

Taking baryons to couple to the metric $g_{\mu\nu}$ and dark matter to couple to a conformally transformed metric $\exp(2\beta\varphi/M_{\rm Pl})g_{\mu\nu}$, with $M_{\rm Pl}=(8\pi G_N)^{-1/2}$ the reduced Planck mass, one has that $Q = (\beta/M_{\rm Pl})\,\dot\varphi\,\rho_{\rm DM}$. This is coupled quintessence \cite{Amendola2000}, the standard realisation of interacting dark energy, in which a rolling quintessence field $\varphi$ exchanges energy with the dark matter at a rate set by the constant coupling $\beta$.
In other formulations --- such as that of an interacting dark-matter fermion \cite{Farrar:2003uw}, or of a pseudo-Nambu--Goldstone field interacting with a ``dark'' baryon \cite{Khoury:2025txd} --- the bare dark-matter mass, $\bar{m}_{\rm DM}$, is instead modified so that it becomes field dependent, $m_{\rm DM}(\varphi)=\bar{m}_{\rm DM}\exp(\beta\varphi/M_{\rm Pl})$, and the resulting dark-matter energy density is $\rho_{\rm DM}=m_{\rm DM}(\varphi)n_{\rm DM}$ where $n_{\rm DM}\propto a^{-3}$.

In these scenarios the cold dark matter particles feel a fifth force of strength $\Delta G/G_N=2\beta^2$ relative to Newtonian gravity, described by the Yukawa potential
\begin{equation}\label{eq:yukawa}
\begin{aligned}
  V_{\rm DM}(r)&=-\frac{G_N {\bar m}_{\rm DM}^2}{r}
    \left[1+\frac{\Delta G}{G_N}\,e^{-r/\lambda_C}\right],\\
  \lambda_C&=\frac{\hbar}{m_\varphi c}
    =1\,\mathrm{Mpc}\left(\frac{m_\varphi}{6.4\times10^{-30}\,\mathrm{eV}}\right)^{-1},
\end{aligned}
\end{equation}
where $m_\varphi$ is the effective mass of the dark-energy scalar field. $\Delta G/G_N$ is positive for either sign of $\beta$, so the force is attractive and the data constrain $|\beta|$. 
A scalar rolling today has $m_\varphi\sim H_0$, hence $\lambda_C\simeq c/H_0\simeq4.4$\,Gpc, nearly three orders of magnitude above the largest range the data constrain directly. (A dictionary between $\beta$ and the other couplings used in the literature is given in Appendix~\ref{app:conv}.)

\section{Data}\label{sec:data}

We use the 4345 edge-on Sloan Digital Sky Survey disks of \DF, selected from the NASA-Sloan Atlas (NSA) by S\'ersic $r$-band absolute magnitude $M_r<-17.7$, distance $D<250$\,Mpc and S\'ersic axis ratio $b/a=0.15$, the floor returned by the profile fits. The axis-ratio cut selects thin disks seen edge-on, and the magnitude cut keeps the galaxies within the well-measured part of the luminosity function on which the halo model of Sec.~\ref{sec:inputs} is calibrated.

We measure each galaxy's warp from its NSA image in the SDSS $r$ band. The major axis $x$ and the perpendicular sky direction $z$ are set by the S\'ersic position angle --- which also fixes the direction along which the forward model of Sec.~\ref{sec:model} projects the fifth-force field --- and $R_{\rm eff}$ is the S\'ersic half-light radius along the major axis. In each pixel column within $|x|<L_x\equiv3R_{\rm eff}$ the disk midline $z(x)$ is the intensity-weighted mean height of the light within $|z|<3(b/a)R_{\rm eff}=0.45R_{\rm eff}$ \cite{DesmondWarps2018,DesmondFerreira2020}. A uniform displacement of the midline is not observable, so we subtract its mean, $z'=z-\langle z\rangle$, and summarise its shape by the dimensionless statistic
\begin{equation}\label{eq:w1def}
  w_1\equiv\frac{1}{L_x^{3}}\int_{-L_x}^{L_x}\!|x|\,z'(x)\,dx ,
\end{equation}
the $|x|$-weighted mean of the midline in units of $L_x$. The weight is even in $x$, so a tilt or the S-shaped (integral-sign) warp common in disk galaxies \cite{SparkeCasertano1988}, both odd in $x$, give $w_1=0$. A U-shaped warp, in which both halves of the disk bend the same way, gives a $w_1$ that grows with the depth of the U and whose sign records its direction on the sky. $w_1$ is therefore a measure of the overall amplitude of a U-shaped warp, the shape a fifth force produces (Sec.~\ref{sec:warpmodel}).
	
Evaluated directly, Eq.~\eqref{eq:w1def} weights every column equally, so its noise is dominated by the faint outer columns, where the midline is poorly determined. We therefore estimate $w_1$ with a matched filter, fitting $z(x)=a+bx+K|x|^{\gamma}$ to the column midlines with weights set by their photon noise and converting $K$ to $w_1$ by the relation derived in Sec.~\ref{sec:warpmodel}. The offset $a$ and tilt $b$ absorb any displacement and S-shaped warp, and $|x|^{\gamma}$, with $\gamma=0.5$, is the shape a fifth force produces in a disk without self-gravity (Sec.~\ref{sec:warpmodel}).
On noiseless synthetic images of a disk warped as $|x|^{0.5}$, at five position angles on the sky, the matched filter underestimates $w_1$ by 1--7\%, depending on whether its weights assume source- or sky-dominated noise; this is included in the factor $S_i$ of Sec.~\ref{sec:inputs}, which is calibrated with the same estimator. The matched filter returns zero for tilted, S-warped and flat disks, and on the survey images its noise is
lower by a median factor of $100$ than that of the direct sum used in \cite{DesmondWarps2018}, whose noise would otherwise significantly exceed the scatter of the warps across the sample.
The same measurement in the $i$ band, less affected by dust, provides a check (Sec.~\ref{sec:systests}).

Of the 4345 galaxies, 4189 have usable images, and 4011 of these are resolved in the sense that the window $|z|<0.45R_{\rm eff}$ spans at least three pixel rows. We remove 38 whose S\'ersic and elliptical Petrosian half-light radii differ by more than a factor of three, the signature of a failed S\'ersic fit, since the forward model uses the S\'ersic radius and position angle. Removing also the five of the seven catalogue artefacts of Sec.~\ref{sec:inputs} that survive these cuts leaves 3968 galaxies. Each measurement has a statistical uncertainty from the photon and read noise of its image, propagated through the fit, with a median of $1.1\times10^{-4}$. The measured values scatter across the sample by $1.4\times10^{-3}$ ($1.48$ times the median absolute deviation), so this scatter is almost entirely astrophysical. The $r$- and $i$-band values have a Spearman correlation of $0.80$. We apply no cut on the measured warp, since the likelihood of Sec.~\ref{sec:inference} accommodates outlying values. These measurements replace the warp amplitudes tabulated in \DF.

\section{Model and inference}\label{sec:model}

\subsection{The warp induced by a dark-sector force}\label{sec:warpmodel}

The Yukawa potential of Eq.~\eqref{eq:yukawa} gives each dark-matter particle an additional acceleration $(\Delta G/G_N)\,\vec{a}_5$, where
\begin{equation}\label{eq:a5}
\begin{aligned}
  \vec{a}_5(\vec{x})&=G_N\!\int\! d^3x'\,\rho_{\rm DM}(\vec{x}')\,
    \frac{\vec{x}'-\vec{x}}{|\vec{x}'-\vec{x}|^{3}}\\
  &\quad\times\left(1+\frac{|\vec{x}'-\vec{x}|}{\lambda_C}\right)
    e^{-|\vec{x}'-\vec{x}|/\lambda_C}
\end{aligned}
\end{equation}
is the fifth-force field per unit coupling sourced by the surrounding dark matter. Stars and gas are uncoupled and do not feel it. A galaxy's halo is therefore accelerated relative to its disk, which, being bound to the halo, settles where the halo's gravity supplies the acceleration the stars are missing, offset from the halo centre along $\vec{a}_5$ (Fig.~\ref{fig:schematic}). We treat this configuration as an equilibrium: a disk settles in a halo dynamical time, $\sim10^{8}$\,yr, two orders of magnitude below the Hubble time over which $\vec{a}_5$ evolves, so we take the force to be the static Yukawa exchange of Eq.~\eqref{eq:yukawa}.\footnote{A rolling scalar adds a second, velocity-dependent acceleration on dark matter which the stars again do not feel, and which therefore warps the disk by the same mechanism. We neglect this here, showing in Sec.~\ref{sec:caveats} that this is conservative.}

The offset itself cannot be seen, but its variation across the disk can. A star at radius $r$ displaced by a height $z$ from the halo's midplane is pulled back by the component of the halo's gravity along $z$, $G_N M_{\rm halo}(<r)\,z/r^3=\tfrac{4\pi}{3}G_N\bar\rho(<r)\,z$, with $\bar\rho(<r)$ the mean density enclosed within $r$. This falls outwards in any halo whose density decreases with radius, so the outer stars must be displaced further than the inner ones to receive the same pull,
 and the disk bends into a U whose arms curve away from $\vec{a}_5$. Balancing, at each cylindrical radius, the halo's gravity perpendicular to the disk against the acceleration that the stars do not feel gives the shape of the midline,
\begin{equation}\label{eq:balance}
\begin{aligned}
  \frac{\Delta G}{G_N}\,a_{5,z}&=-\frac{G_N M_{\rm halo}(<r)}{r^{2}}\,\frac{z}{r},\\
  \Longrightarrow\quad z(x)&=-\frac{\Delta G}{G_N}\,
    \frac{a_{5,z}\,|x|^{3}}{G_N M_{\rm halo}(<|x|)} ,
\end{aligned}
\end{equation}
with $x$ along the major axis and $M_{\rm halo}(<r)$ the halo mass enclosed within $r$. The gas, being uncoupled, moves with the stars, so among the coupled components only the halo restores the disk. We assume the displacement to be small compared with the disk radius, so that the response is linear in the force.

For an inner halo profile $\rho=\rho_s(r/r_s)^{-\gamma}$, with $r_s$ the halo's scale radius and $\rho_s$ the density there, Eq.~\eqref{eq:balance} integrates to a power law, $z(x)=K|x|^{\gamma}$ with $K\equiv-(a_{5,z}/G_N)(\Delta G/G_N)(3-\gamma)/(4\pi\rho_s r_s^{\gamma})$, whose warp amplitude, Eq.~\eqref{eq:w1def}, is $w_1=\gamma K L_x^{\gamma-1}/[(\gamma+1)(\gamma+2)]$. Writing $a_{5,\perp}$ for the component of $\vec{a}_5$ perpendicular to the disk major axis on the sky, which for a disk seen edge-on is $a_{5,z}$, substitution for $K$ gives \cite{DesmondWarps2018}
\begin{equation}\label{eq:w1}
  w_1=-\frac{\gamma(3-\gamma)}{(\gamma+1)(\gamma+2)}\,
  \frac{\Delta G}{G_N}\,\frac{a_{5,\perp}}{4\pi G_N\rho_s}\,\frac{1}{L_x}
  \left(\frac{L_x}{r_s}\right)^{\!\gamma} .
\end{equation}
The amplitude is exactly linear in $\Delta G/G_N$,
every other factor being fixed by the observed galaxy and its environment.
Eq.~\eqref{eq:w1} omits two effects that reduce the observed warp at fixed coupling. First, the disk's own gravity resists the bend, since annuli displaced by different heights attract one another. This correction can be significant: in the median galaxy of our sample the total stellar mass $M_\star$ is 11 times the mass $M_{\rm halo}(<R_{\rm eff})$ of its abundance-matched halo (Sec.~\ref{sec:inputs}) within $R_{\rm eff}$, and even the stellar mass within $R_{\rm eff}$, $M_\star(<R_{\rm eff})\simeq0.24M_\star$ for the exponential disk of Sec.~\ref{sec:inputs}, is three times $M_{\rm halo}(<R_{\rm eff})$. We therefore solve for the equilibrium midline of an exponential disk of annuli, adding to Eq.~\eqref{eq:balance} the vertical force on each annulus from the others, to linear order in their relative displacements and softened over the disk thickness. The ratio $M_{\rm halo}(<R_{\rm eff})/M_\star$ then sets how far the disk stiffens: a disk dominated by its halo follows Eq.~\eqref{eq:w1}, while one dominated by its own mass bends little in its inner parts. Second, a disk seen edge-on presents at projected position $x$ the light of every annulus of radius $R\ge|x|$, so the observed midline is the light-weighted mean of $z(R)$ along the line of sight, which flattens the U. Both effects preserve the linearity in $\Delta G/G_N$, and we fold them into the factor $S_i$ of Sec.~\ref{sec:inputs}. Finally, the force acts on the halo only if the galaxy is unscreened, which we assume; this is discussed further in Sec.~\ref{sec:screening}.

\subsection{Inputs to the forward model}\label{sec:inputs}

Evaluating Eq.~\eqref{eq:w1} requires, for each galaxy, $\vec{a}_5$ and the halo parameters $\rho_s$, $r_s$ and $\gamma$. We take $\vec{a}_5$, $\rho_s$ and $r_s$ from \DF: $\vec{a}_5$ from a model of the mass distribution of the local Universe, and $\rho_s$ and $r_s$ from abundance matching, which also supplies the halos in that model (see below).

We evaluate Eq.~\eqref{eq:a5} by summing over the mass within $4\lambda_C$ of the galaxy, beyond which the Yukawa factor $(1+r/\lambda_C)\,e^{-r/\lambda_C}$ falls below $0.1$. This mass has three components.
The first comprises the halos of the galaxies in the NSA, assigned by abundance matching (see below) and given Navarro--Frenk--White (NFW) profiles. The second comprises the halos of galaxies too faint for the survey, whose contribution is predicted statistically from that of the visible ones by regressors trained on the DarkSky-400 $N$-body simulation \cite{Skillman2014}. The third is the mass outside halos, from the \texttt{BORG} reconstruction of the local Universe \cite{JascheLavaux2019}, which fits a particle-mesh model of structure formation from Gaussian initial conditions to the 2M++ galaxy compilation \cite{LavauxHudson2011}. This yields a posterior over the present-day matter density, resolved to $\sim$3\,Mpc, out to 290\,Mpc, beyond which 2M++ becomes incomplete. We use 10 of its samples, from each of which the mass already assigned to halos is subtracted. Restricting the sample to $D<250$\,Mpc keeps every source within $4\lambda_C$ of a test galaxy inside this volume for $\lambda_C\le8$\,Mpc.

Abundance matching assigns halos to galaxies by rank: galaxies ordered by luminosity are matched, with scatter, to simulated halos ordered by a mass proxy, so that their number densities agree \cite{Behroozi2010,Moster2013}. \DF\ use the model of \cite{Lehmann2017}, whose proxy interpolates between a halo's peak mass and peak maximal circular velocity, applied to the DarkSky-400 halos with the SDSS $r$-band luminosity function to make 200 mock catalogues, each with the model's two parameters (the halo proxy parameter, governing the interpolation, and the scatter) drawn from their posterior from SDSS galaxy clustering. In each Monte Carlo realisation a galaxy's true magnitude is drawn from its measurement uncertainty and the galaxy is assigned the halo of closest magnitude in a mock catalogue. This gives the source halos of $\vec{a}_5$ their masses and concentrations $c_{\rm vir}$, and each test galaxy the scale radius $r_s=r_{\rm vir}/c_{\rm vir}$ of its halo and the NFW density there, $\rho_s$. Their distributions over the realisations, which reflect the uncertainties in both the galaxy's luminosity and the galaxy--halo connection, are propagated into the predicted signal.
 Galaxy formation alters the inner slope, adiabatic contraction steepening it and stellar feedback flattening it.
 The profiles inferred dynamically for many galaxies are shallower than the NFW $\gamma=1$ \cite{deBlok2010,Oh2015}, so we follow \DF\ in taking $\gamma=0.5$ as fiducial and test $\gamma=1$ in Sec.~\ref{sec:systests}. A fully cored halo, $\gamma\to0$, would produce no U-shaped warp, since its restoring force per unit displacement is then independent of radius.

Each galaxy's $w_1$ per unit $\Delta G/G_N$ from Eq.~\eqref{eq:w1} is a distribution, obtained by Monte Carlo propagation of the uncertainties in the galaxy and halo properties and in the density field. For seven galaxies the mean of this distribution lies between 10 and 950, more than ten times the 99th percentile of the sample, $0.8$.
 Their Monte Carlo distributions are narrow, so these values are not chance draws but come from the input catalogues. For four of them the source catalogue holds a second entry within $1.2$\,kpc (a duplicate or close companion that the sum treats as a separate halo), two have effective radii of $0.2$\,kpc (more than an order of magnitude below the sample median of $5$\,kpc) and one lies $0.2$\,Mpc from a source assigned a halo mass of $2\times10^{15}\,M_\odot$. We remove them;
if kept they would dominate the bound, tightening it by a factor 1.9.

The warp that the measurement of Sec.~\ref{sec:data} returns differs from the $w_1$ of Eq.~\eqref{eq:w1}, both because the disk's self-gravity and edge-on projection change the shape of the midline (Sec.~\ref{sec:warpmodel}) and because seeing, pixelisation and the weighting of the matched filter change how that shape is measured. We absorb all of these into a factor $S_i$ for each galaxy, the ratio of the warp the measurement would return to the $w_1$ of Eq.~\eqref{eq:w1}, which we calibrate on synthetic images.

To do this we build an image of each galaxy's disk with its own half-light radius $R_{\rm eff}$, position angle, pixel grid, flux and noise, convolved with the seeing of its SDSS field (median full width at half maximum $1.3$\,arcsec). The disk is exponential with scale length $R_{\rm eff}/0.94$,
with a ${\rm sech}^2$ vertical profile of scale height $0.15R_{\rm eff}$ and the stellar mass $M_\star$ from the S\'ersic photometry of the NSA. We warp it by the midline of Sec.~\ref{sec:warpmodel}, computed for the galaxy's own $M_{\rm halo}(<R_{\rm eff})/M_\star$ with self-gravity and projection included, and apply the matched filter to the warped and unwarped images. $S_i$ is the difference between the two measurements divided by the $w_1$ of Eq.~\eqref{eq:w1}.

$S_i$ has a median of $0.27$, with 16th and 84th percentiles of $0.05$ and $0.39$. Self-gravity dominates this suppression: the median $S_i$ rises from $0.03$ for galaxies with $M_{\rm halo}(<R_{\rm eff})/M_\star<0.03$ to $0.45$ for those with $M_{\rm halo}(<R_{\rm eff})/M_\star>0.3$, whereas projection alone gives $S_i\simeq0.58$ and seeing lowers $S_i$ by only $2\%$ in the median, because a self-gravitating disk bends mostly in its outer parts.

We call $S_i w_1$ per unit $\Delta G/G_N$, the warp the measurement would return, the predicted signal. At the conservative bound of Eq.~\eqref{eq:headline} the 99th-percentile galaxy has a predicted warp of $w_1=3.2\times10^{-4}$, corresponding to a midline displacement of about $35$\,pc at $L_x=15$\,kpc, or $2\times10^{-3}$ of its radius, so the terms of relative order $(z/x)^2$ dropped from Eq.~\eqref{eq:balance} are negligible.

\subsection{Likelihood}\label{sec:inference}

We fit each galaxy's distribution of $w_1$ per unit $\Delta G/G_N$ from Eq.~\eqref{eq:w1} with a mixture of $n_i\le15$ Gaussians of weights $W_{ik}$, means $M_{ik}$ and variances $C_{ik}$, with $n_i$ set per galaxy by the Bayesian information criterion. Bars, tides and interactions occasionally produce warps far larger than those of the bulk of the sample, so we compare the predicted signal with the measurement through a Student-$t$ distribution of scale $\sigma_{w_1}$ and $\nu$ degrees of freedom, both free parameters of the inference; the heavy tails of the Student-$t$ distribution accommodate such galaxies without a cut on the measured warp such as was used in \DF. This yields
\begin{equation}\label{eq:like_mixture}
\begin{aligned}
  \mathcal{L}\!\left(\tfrac{\Delta G}{G_N},\sigma_{w_1},\nu\right)
  &=\prod_{i}\sum_{k=1}^{n_i} W_{ik}\,
  t_\nu\!\left(w_{1,i}^{\rm obs}\;\Big|\;\tfrac{\Delta G}{G_N}S_iM_{ik},\,s_{ik}\right),\\
  s_{ik}^2&=\sigma_{w_1}^{2}+e_i^2+\left(\tfrac{\Delta G}{G_N}\right)^{2}S_i^2C_{ik},
\end{aligned}
\end{equation}
with $w_{1,i}^{\rm obs}$ and $e_i$ the measured warp of galaxy $i$ and its uncertainty, and $t_\nu(x\,|\,\mu,s)$ the Student-$t$ density of location $\mu$ and scale $s$. The coupling scales every component mean and enters the variance in quadrature, so the predicted scatter grows with the predicted signal. $\sigma_{w_1}$ and $\nu$ describe the astrophysical scatter in warp amplitude. We take flat priors on $\Delta G/G_N\ge0$, $\sigma_{w_1}\in[2\times10^{-4},3\times10^{-3}]$ and $\ln\nu\in[\ln0.5,\ln100]$, and evaluate the posterior of $\Delta G/G_N$ on a fixed grid, integrating out $\sigma_{w_1}$ and $\nu$ at each grid point by the Laplace approximation (Appendix~\ref{app:neglected}).

\subsection{Systematics tests}\label{sec:systests}

We test the sensitivity of the bound to variations in the model assumptions, as listed in Table~\ref{tab:bounds}(b). One variation replaces the $r$-band warps by the $i$-band ones. Two excise the 1\% or 5\% of galaxies with the largest predicted signal, ranked by $S_i$ times the mixture-weighted mean of $|M_{ik}|$, to test how far the bound rests on the disks the force would bend most. One replaces the noise model by a mixture of a Gaussian and a Student-$t$ distribution of free relative weight, both centred on the predicted signal, and two remove the galaxies whose measured warp exceeds three or five times the scatter of Sec.~\ref{sec:data}.
For these we use a truncated likelihood to account for this selection effect: each remaining galaxy's density is renormalised to the range of warps that passes the cut, so that the cut by itself does not favour models predicting small warps.
The other variations change the forward model: the disk scale height in $S_i$ ($0.1R_{\rm eff}$ or $0.2R_{\rm eff}$), the stellar masses ($\pm0.2$\,dex), the disk scale length ($R_{\rm eff}/1.68$, appropriate to a face-on disk), the inner halo slope ($\gamma=1$), a self-gravitating gas disk that moves with the stars (an exponential of twice the stellar scale length, with the median gas-to-stellar mass ratio of \cite{Catinella2018} at the galaxy's stellar mass) and a $30\%$ fractional uncertainty on every predicted signal, added in quadrature to $S_i^2C_{ik}$.

The final variation, masking foreground stars, is a more significant departure from the fiducial data reduction.
A star inside the window of Sec.~\ref{sec:data} adds light at one height in the columns it overlaps and so shifts their midlines. The NSA pipeline removes sources consistent with the point-spread function when it deblends each galaxy \cite{Blanton2011}, but not always completely. We therefore identify as stars the Gaia DR3 sources \cite{GaiaMission,GaiaDR3} near each galaxy whose parallax or proper motion is detected at more than $5\sigma$, and measure how much of each star's light remains in the image. To do so we fit the image within twice the seeing full width at half maximum of the star's position with a point-source model of the star's Gaia flux, converted to the SDSS band and multiplied by a free amplitude $A$, plus a plane for the light of the galaxy and sky beneath it. $A=0$ means that the deblending removed the star entirely and $A=1$ that it left the star in place. We find $|A|<0.05$ for 96\% of stars, but $A>0.5$ for 1.7\%. Taking each star's residual light to be its point-source model times $|A|+2\sigma_A$, with $\sigma_A$ the uncertainty in $A$, we remove from the fit every column whose midline that light would shift by more than $0.2$ times the statistical uncertainty of the midline, and recompute $S_i$ with the same columns removed. This removes a median $6\%$ of the fit window in 408 galaxies of the fiducial sample and the whole window of one compact galaxy.

Each variation is an alternative to one choice in the fiducial analysis, so we form our conservative bound by multiplying the fiducial bound by the largest factor by which any variation weakens it, evaluated separately at each $\lambda_C$. The excisions require a correction to this: removing the galaxies with the largest predicted signal discards the most informative data and so weakens the bound even when the model correctly describes all the data. For each excision we therefore use the ratio of the excised to the full-sample bound divided by the median of the same ratio over mock data sets with no fifth force (Appendix~\ref{app:neglected}), floored at one, so that only a weakening beyond that expected from the loss of information as the sample size is reduced counts as a systematic. Finally, we multiply by $1.14$ to account for the Monte Carlo noise in the predicted signals. This factor is one plus the rms fractional scatter of the fiducial bound about a smooth function of $\lambda_C$ across the twenty force ranges, whose predicted signals come from independent Monte Carlo runs. This is not a systematic \textit{per se}, and can be arbitrarily reduced by increasing the number of Monte Carlo draws, but effectively acts as an unmodelled uncertainty in our fiducial analysis. The screened model also listed in Table~\ref{tab:bounds}(b) is discussed in Sec.~\ref{sec:screening}.

\begin{table*}
\centering
\caption{Constraints on the strength $\Delta G/G_N$ of a dark-sector fifth force from the warping of stellar disks.
(a) The 95\% bound at three of the force ranges of Fig.~\ref{fig:exclusion} and at longer range (Sec.~\ref{sec:bound}); all twenty ranges are available in the public release. Conservative values multiply the fiducial ones by the largest factor from the variations of panel (b), evaluated separately at each range, and by $1.14$ for Monte Carlo noise (Sec.~\ref{sec:systests}). (b) The bound at $\lambda_C=8$\,Mpc under each variation (Sec.~\ref{sec:systests}).
For the excisions the number in parentheses is the ratio of the excised to the full-sample bound divided by its median over 40 mock data sets with no fifth force (Appendix~\ref{app:neglected}), floored at one, which is the factor entering the conservative bound.
 (c) Published claims and limits for a long-range dark-sector force. Significances divide each central value by the quadrature sum of its own uncertainty and ours, the latter from a half-Gaussian whose 95\% point is our conservative bound.
This involves extrapolating the Gaussian approximation into its tail, so these values are indicative only. A source that fits $\tilde{c}$ has an interval skewed by the transformation to $\Delta G/G_N$, which we linearise: for \cite{Bedroya2025} this lowers the significance from $2.8\sigma$ in $\tilde{c}$ to the $2.3\sigma$ quoted. Factors compare with our conservative bound. The limit of \cite{Bottaro2025} at $8$\,Mpc is interpolated from their Fig.~1 (Sec.~\ref{sec:implications}).}
\label{tab:bounds}
\renewcommand{\arraystretch}{0.92}
\begin{tabular}{lcc}
\hline\hline
\multicolumn{3}{l}{\textbf{(a) Bound vs.\ force range} (all galaxies unscreened)}\\
\hline
$\lambda_C$ [Mpc] & fiducial & conservative\\
\hline
0.4 & $4.6\times10^{-3}$ & $1.3\times10^{-2}$\\
2.0 & $1.6\times10^{-3}$ & $3.8\times10^{-3}$\\
8.0 & $7.7\times10^{-4}$ & $1.7\times10^{-3}$\\
$>8$, incl.\ $m_\varphi\!\to\!0$ & $<7.7\times10^{-4}$ & $<1.7\times10^{-3}$\\
\hline
\multicolumn{3}{l}{\textbf{(b) Systematic variations} (95\%, $\lambda_C=8$\,Mpc)}\\
\hline
Variation & $\Delta G/G_N$ & ratio\\
\hline
Fiducial ($r$ band, $N=3968$)              & $7.7\times10^{-4}$ & 1.00\\
$i$ band instead of $r$ band ($N=3963$)    & $7.2\times10^{-4}$ & 0.95\\
Excise top 1\% by predicted signal         & $1.6\times10^{-3}$ & 2.03 (1.24)\\
Excise top 5\% by predicted signal         & $1.6\times10^{-3}$ & 2.13 (1.00)\\
Gaussian core + Student-$t$ noise          & $6.8\times10^{-4}$ & 0.88\\
Remove $|w_1|>3\sigma$, truncated likelihood ($N=3890$) & $1.3\times10^{-3}$ & 1.74\\
Remove $|w_1|>5\sigma$, truncated likelihood ($N=3956$) & $9.2\times10^{-4}$ & 1.20\\
Scale height $0.1R_{\rm eff}$              & $7.5\times10^{-4}$ & 0.98\\
Scale height $0.2R_{\rm eff}$              & $8.2\times10^{-4}$ & 1.08\\
Scale length $R_{\rm eff}/1.68$ (face-on)  & $1.4\times10^{-3}$ & 1.88\\
Stellar mass $+0.2$\,dex                   & $9.6\times10^{-4}$ & 1.26\\
Stellar mass $-0.2$\,dex                   & $6.3\times10^{-4}$ & 0.83\\
Inner halo slope $\gamma=1$                & $6.0\times10^{-4}$ & 0.79\\
Self-gravitating gas disk                  & $9.1\times10^{-4}$ & 1.19\\
$30\%$ forward-model uncertainty           & $9.3\times10^{-4}$ & 1.21\\
Foreground stars masked ($N=3967$) & $7.3\times10^{-4}$ & 0.95\\
Screening at $\chi=\Phi^{\rm LSS}_{\rm rms}/c^2$, prob.\ weighted & $1.3\times10^{-3}$ & 1.66\\
\hline
\textbf{Conservative bound}                  & $\mathbf{1.7\times10^{-3}}$ & 2.15\\
\hline
\multicolumn{3}{l}{\textbf{(c) Published claims and limits} ($\lambda_C\to\infty$ unless stated)}\\
\hline
Source & $\Delta G/G_N$ & disfavoured at\\
\hline
DESI DR2 BAO~\cite{Clini2026}        & $0.0037\pm0.0015$ & $2.2\sigma$\\
DESI DR1 FS~\cite{Clini2026}         & $0.0069\pm0.0021$ & $3.1\sigma$\\
DESI DR1 FS + DR2 BAO~\cite{Clini2026}    & $0.0043\pm0.0014$ & $2.6\sigma$\\
DESI BAO~\cite{Bottaro2025}        & $0.0040\pm0.0020$ & $1.8\sigma$\\
DESI DR2 + SNe~\cite{Bedroya2025}  & $0.0050\pm0.0020$ & $2.3\sigma$\\
Coupled DE, DESI ($|\beta|\simeq0.03$)~\cite{CoupledDESI2026} & $1.8\times10^{-3}$ & $2.1\sigma$\\
Swampland, $\tilde{c}=0.30$~\cite{Agrawal2019} & $0.18$ & $>100\sigma$\\
\hline
\multicolumn{3}{l}{\emph{Upper limits; final column gives the factor by which this work is stronger}}\\
Planck + BOSS FS ($\lambda_C=8$\,Mpc)~\cite{Bottaro2025} & $<9.8\times10^{-4}$ & $0.6\times$\\
Planck + BOSS FS~\cite{Bottaro2024} & $<4.8\times10^{-3}$ & $2.9\times$\\
CMB + BAO~\cite{Archidiacono2022}   & $<5.4\times10^{-3}$ & $3.3\times$\\
CMB + LSS + SNe ($\lambda_C\simeq10$\,Mpc)~\cite{Bean2008} & $<0.05$ & $30\times$\\
Euler equation~\cite{Bonvin2025}    & $<0.07$ & $43\times$\\
Sagittarius stream~\cite{Kesden2006}        & $<0.1$  & $61\times$\\
Fornax~\cite{Zamlung2026}           & $<0.6$  & $360\times$\\
\hline\hline
\end{tabular}
\end{table*}

\section{Results -- bounds on \texorpdfstring{$\Delta G/G_N$}{DG/GN}}\label{sec:bound}
\begin{figure}[tb]
  \centering
  \includegraphics[width=\columnwidth]{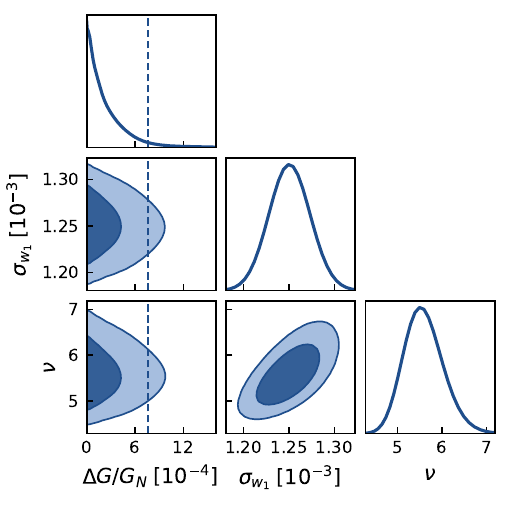}
  \caption{Posterior of $\Delta G/G_N$ and the noise parameters $\sigma_{w_1}$ and $\nu$ of Eq.~\eqref{eq:like_mixture} at $\lambda_C=8$\,Mpc.
Contours enclose 68\% and 95\% of the two-dimensional marginal posteriors, and the dashed line marks the 95\% upper limit on $\Delta G/G_N$.
}
  \label{fig:corner}
\end{figure}

Marginalising Eq.~\eqref{eq:like_mixture} over $\sigma_{w_1}$ and $\nu$, the warps of 3968 edge-on disks bound a dark-sector fifth force at $\Delta G/G_N<7.7\times10^{-4}$ (95\% credibility) and $<2.8\times10^{-4}$ (68\%) at $\lambda_C=8$\,Mpc, the largest fifth-force range investigated in \DF. The noise posterior peaks at $\sigma_{w_1}=1.25\times10^{-3}$ and $\nu=5.6$, accounting for essentially the entire width of the measured $w_1$ distribution, and is almost independent of $\Delta G/G_N$ (Fig.~\ref{fig:corner}). This is the desired behaviour: warps are astrophysical, and the fifth force is sought only in their galaxy-by-galaxy correlation with $\vec{a}_5$. Allowing $\Delta G/G_N$ to take either sign gives $(-4.6\pm3.2)\times10^{-4}$ in the $r$ band and $(-6.2\pm4.3)\times10^{-4}$ in the $i$ band, so the measured warps lean slightly against the predicted direction.
Almost all of this lean comes from a single galaxy whose image retains a bright foreground star that the NSA deblending left in place ($A\simeq1$). Our model variation masking foreground stars (Sec.~\ref{sec:systests}), which removes the columns of this galaxy's window that the star contaminates, reduces the two values to $(-1.7\pm3.1)\times10^{-4}$ and $(-1.8\pm4.1)\times10^{-4}$. The bound is somewhat tighter than is typical for mock data generated with no fifth force, for which the median 95\% limit is $(8$--$11)\times10^{-4}$ (Appendix~\ref{app:neglected}).

The variations of Sec.~\ref{sec:systests} return bounds between $0.79$ and $2.13$ times the fiducial value at $8$\,Mpc (Table~\ref{tab:bounds}(b)). The excisions of the top 1\% and 5\% weaken it by factors $2.0$ and $2.1$ (cf. median factors of $1.6$ and $2.7$ over mock data sets with no fifth force), so they enter the conservative bound as factors $1.24$ and $1.00$. Masking foreground stars changes the fiducial bound by a factor $0.95$ at $8$\,Mpc and by $0.70$--$1.02$ across the twenty ranges. The largest change, at $0.4$\,Mpc, comes from the galaxy whose image retains a star, and rebuilding the conservative bound from variations evaluated on the masked warps leaves it unchanged at $8$\,Mpc. Applying the procedure of Sec.~\ref{sec:systests}, the largest factor at $8$\,Mpc is $1.88$, from replacing the edge-on disk scale length $R_{\rm eff}/0.94$ used to compute $S_i$ by the face-on $R_{\rm eff}/1.68$. Across the twenty ranges the largest factor lies between $1.6$ and $2.4$, set either by this variation or by the removal of galaxies whose measured warp exceeds three times the scatter. Multiplying the fiducial bound by this factor and by $1.14$ gives our conservative bound
\begin{equation}\label{eq:headline}
  \frac{\Delta G}{G_N}<1.7\times10^{-3}\quad(95\%)\quad\text{at}\quad \lambda_C=8\,\mathrm{Mpc},
\end{equation}
or equivalently $|\beta|<0.029$ for the conformal coupling $\beta$ of Sec.~\ref{sec:coupling}. Note that this should not be considered a credible interval in the normal statistical sense, deriving from a single model; rather it is an envelope over multiple possible models that is 
robust to every variation in Table~\ref{tab:bounds}(b).

The 5\% of galaxies with the largest predicted signal, which set the bound, are not anomalous but differ from the bulk of the sample in two respects. First, they lie in fifth-force fields stronger by a factor 4.4 in the median. Second, their disks are less massive, with a median stellar mass $2.6$ times lower in halos of the same median density $\rho_s$, so that the halo supplies a larger share of the restoring force within $R_{\rm eff}$ (median $M_{\rm halo}(<R_{\rm eff})/M_\star$ of $0.18$, versus $0.09$ for the rest of the sample) and self-gravity suppresses their warps less (median $S_i$ of $0.35$, versus $0.26$).

\begin{figure*}[tb]
  \centering
  \includegraphics[width=\textwidth]{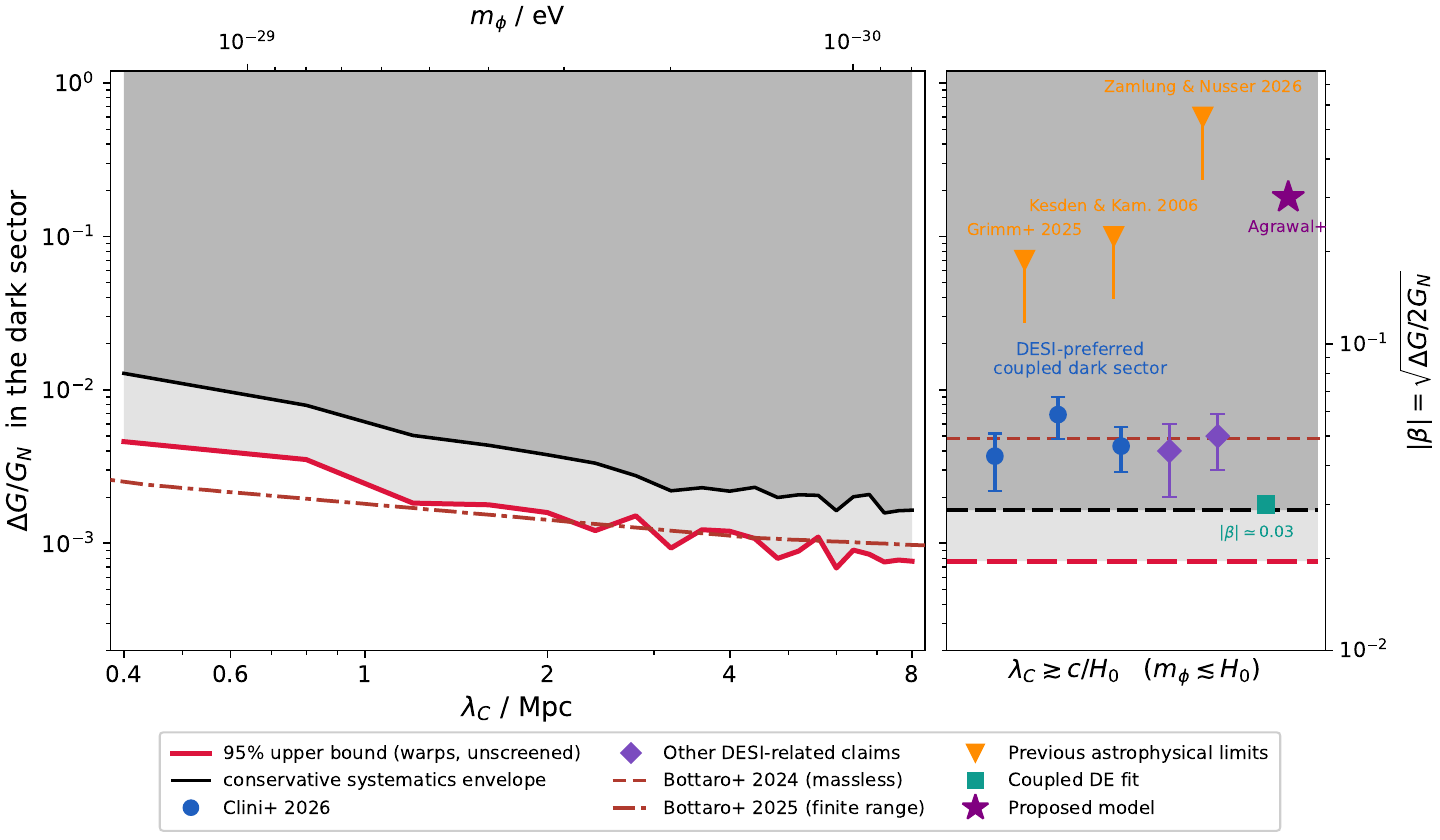}
  \caption{Exclusion on the strength $\Delta G/G_N$ of a fifth force acting on dark matter but not
  stars, against the force range $\lambda_C$ (top axis: scalar mass; right axis:
  $|\beta|=\sqrt{\Delta G/2G_N}$). {\textbf{Left:} the range the warp data constrain
  directly.} \textbf{Right:} the effectively infinite-range regime, where the dashed lines conservatively continue the $8$\,Mpc bound (horizontal positions in that panel are arbitrary). Solid red is the fiducial 95\% limit and solid black the corresponding conservative limit, inflated for systematic uncertainties.
The shading marks the
  excluded region, dark above the inflated bound and light between it and the fiducial limit.
  Blue circles are the three data combinations of \cite{Clini2026} and purple diamonds the
  inferences of \cite{Bottaro2025,Bedroya2025} (Table~\ref{tab:bounds}(c)). The teal
  square is a coupled dark energy fit \cite{CoupledDESI2026}, orange triangles are previous astrophysical upper limits,
  the dark-red dashed horizontal line is the 95\% cosmological bound of \cite{Bottaro2024} for $m_\varphi<H_0$, the dark-red dash-dotted line in the left panel is the 95\% cosmological bound of \cite{Bottaro2025} for a scalar with mass, digitised from their Fig.~1 (Sec.~\ref{sec:implications}), and the star is
  the model of \cite{Agrawal2019} at $\tilde{c}=0.30$.
  }
  \label{fig:exclusion}
\end{figure*}

The left panel of Fig.~\ref{fig:exclusion} shows the fiducial 95\% limit (red) over the range $0.4$--$8$\,Mpc that the data constrain directly, together with the conservative limit (black) and the cosmological bound of \cite{Bottaro2025} (dash-dotted, Sec.~\ref{sec:implications}). The limit tightens with range---approximately as $\Delta G/G_N\lesssim7.7\times10^{-4}(\lambda_C/8\,{\rm Mpc})^{-0.6}$---because a longer-ranged force is sourced by more mass and so predicts larger warps. It improves by a factor $6.0$ between $\lambda_C=0.4$ and $8$\,Mpc. The right panel continues the $8$\,Mpc limits to $\lambda_C\gtrsim c/H_0$, which is shown below to be a conservative extrapolation. It also compares to previous literature claims and limits, which we summarise in Table~\ref{tab:bounds}(c) and discuss in Sec.~\ref{sec:implications}.

Given that the density-field reconstruction extends only to $290$\,Mpc, we cannot compute the fifth-force signal directly out to the $\lambda_C\gtrsim4$\,Gpc massless limit relevant for a rolling scalar, and must instead argue from our $8$\,Mpc result. There are two independent arguments for why this should only tighten going out to larger scales. First, an analytic calculation of the isotropic fifth-force field shows that its expected strength grows monotonically as the force gets lighter (longer-ranged), so a lighter scalar sources a larger signal on average, although this alone does not guarantee a tighter bound.
Second, more empirically, we have rerun the forward model at $12$, $16$ and $20$\,Mpc, as far as the reconstruction allows, in pairs of runs that share their random seeds with runs at $8$\,Mpc, and find the bound on $\Delta G/G_N$ to strengthen with range, to $0.76\pm0.05$ of its $8$\,Mpc value at $20$\,Mpc. We therefore adopt the $8$\,Mpc bound as a conservative limit at every longer range, including the massless limit. The full analytic and numerical argument is given in Appendix~\ref{app:massless}.

\section{Screening}\label{sec:screening}

The most obvious way to evade these constraints is by equipping the scalar field with a screening mechanism, causing it to decouple within galaxies while remaining active on cosmological scales. Chameleon and symmetron scalars screen in dense environments according to the gravitational potential, $\Phi$. An object is unscreened where $|\Phi|/c^2$ falls below a self-screening parameter $\chi$ set by the background field value, with $\Phi$ summed over the object itself and its neighbours within $\lambda_C$ \cite{Zhao2011PRL,Cabre2012,BurrageSakstein2018}. Screening the galaxies that set our bound requires $\chi\lesssim6\times10^{-5}$: the deepest potential in the sample is $|\Phi|/c^2=1.3\times10^{-4}$, and the 1\% of galaxies with the largest predicted signal have a median of $3.1\times10^{-5}$, or $5.9\times10^{-5}$ at the upper end of each galaxy's potential interval. The question is whether $\chi$ can be that small while the force remains cosmologically relevant.

For a scalar whose cosmological value tracks the minimum of its effective potential, as in chameleon, symmetron and dilaton models and $f(R)$ gravity, it cannot. Differentiating the minimum condition in time ties the drift of $\chi$ to the scalar's mass, $d\chi/d\ln a=\tfrac92\Omega_m(H\lambda_C/c)^2$ with $\Omega_m$ the matter density parameter, so that today $\chi\gtrsim(H_0\lambda_C/c)^2$ up to a factor of order unity. The same relation underlies the no-go theorem of \cite{WangHuiKhoury2012}, which confines tracking chameleons that screen the Milky Way to $\lambda_C\lesssim1$\,Mpc. In the Hu--Sawicki model of $f(R)$ gravity \cite{HuSawicki2007}, for example, the background scalaron amplitude $f_{R0}$ sets $\chi=\tfrac32 f_{R0}$ and, for the fiducial exponent $n=1$, $\lambda_C\simeq3.2\,\mathrm{Gpc}\,\sqrt{f_{R0}}$, so that
\begin{equation}\label{eq:chilam}
  \chi \simeq \frac{3}{2}\left(\frac{\lambda_C}{3.2\,\mathrm{Gpc}}\right)^{2} .
\end{equation}
Screening our galaxies therefore requires $\lambda_C\lesssim20$\,Mpc for Hu--Sawicki $f(R)$, and $\lambda_C\lesssim(c/H_0)(6\times10^{-5})^{1/2}\simeq35$\,Mpc in general up to the same order-unity factor, ranges with no reach on the scales the cosmological fits measure. Conversely, a scalar rolling today, as the DESI inferences invoke, has $m_\varphi\sim H_0$ and $\lambda_C\gtrsim c/H_0$, for which $\chi$ is of order unity ($3$ from Eq.~\eqref{eq:chilam}), far above the potential of any galaxy or even cluster core (${\sim}10^{-4}$). Every galaxy in the sample is then unscreened and the bound applies at full strength.
For Hu--Sawicki $f(R)$ the short-range window is in any case closed by \DF, who fit the screened model to the warps of these galaxies together with gas--star offsets and achieve constraints at the $f_{R0} < \text{few} \times 10^{-8}$ level.

A model could escape this relation, most simply by not tracking its minimum \cite{WangHuiKhoury2012}, and so combine a long range with a small $\chi$. However, it still could not screen our sample if it is to alter the growth of structure that the full-shape fits measure \cite{Bottaro2024,Clini2026}, since for $\chi$ below the rms potential of the linear density modes, $\Phi^{\rm LSS}_{\rm rms}/c^2=4.2\times10^{-5}$, those modes would be screened too.
At $\chi=4.2\times10^{-5}$, $97\%$ of our galaxies are unscreened, falling to $79\%$ and $70\%$ in the top $5\%$ and $1\%$ by predicted signal, which sit in deeper potentials because a large $\vec{a}_5$ requires a nearby massive structure. To quantify the effect at this $\chi$ we assign each galaxy the probability $f_i$ that it is unscreened, i.e.\ that its potential lies below $\chi$, estimated from the distribution of its potential over the Monte Carlo realisations. Following \DF's treatment of thin-shell screening, galaxy $i$ then contributes to the likelihood $f_i$ times its term in Eq.~\eqref{eq:like_mixture} plus $1-f_i$ times the probability of its observed warp with no fifth force, $t_\nu(w_{1,i}^{\rm obs}\,|\,0,\sqrt{\sigma_{w_1}^2+e_i^2})$. This weakens the $95\%$ bound by a factor $1.66$ (Table~\ref{tab:bounds}(b)), within the factor $2.15$ by which the conservative bound exceeds the fiducial one.
This type of screening therefore cannot evade the test.

There are two kinds of coupling that could. Vainshtein screening
\cite{Vainshtein1972,Kobayashi2019} suppresses the force within
$r_V=(16G_NM/9m_\varphi^2)^{1/3}$ (with $\hbar=c=1$). For $m_\varphi=H_0$ that is 1.2\,Mpc at $10^{12}\,M_\odot$ and
5.5\,Mpc at $10^{14}\,M_\odot$, enclosing the sample and its neighbours. A disformal coupling gives
no force between static configurations \cite{Koivisto2012}. Further investigation of these is left for future work.

\section{Remaining Systematic Uncertainties}\label{sec:caveats}

There are a number of remaining potential systematics. In roughly decreasing order of importance:
\begin{itemize}
\item The halo response to baryons is unmodelled. Adiabatic contraction by the disk raises the inner halo mass, which lowers the warp of Eq.~\eqref{eq:w1} at fixed coupling but also lessens the suppression by the disk's self-gravity (Sec.~\ref{sec:warpmodel}). Our $\gamma=1$ variation, a steeper inner profile with more halo mass inside the disk, tightens the bound by a factor $0.79$, although a contracted halo could produce a different profile shape. Galaxy-by-galaxy mass models would be needed to settle the size of this effect.
\item The noise terms of Eq.~\eqref{eq:like_mixture} treat baryonic warps as independent of the galaxy and its environment, which ram pressure, feedback and tides need not respect. However, the calibration of \cite{Bartlett2021} against Horizon-AGN showed this to be a good approximation in practice for this observable (and for this simulation).
\item The factors $S_i$ rest on an idealised disk: exponential, with a constant ${\rm sech}^2$ scale height, seen through a Gaussian point-spread function and with stellar masses from the NSA photometry. The variations of Table~\ref{tab:bounds}(b) in scale height, scale length, stellar mass and gas content change the bound by up to a factor $1.88$. The gas variation uses median gas fractions over all galaxy types, which may understate those of disk galaxies.
Dust in the midplane of an edge-on disk flattens its light profile, as our face-on scale-length variation does, but is not otherwise modelled.
\item The likelihood of Eq.~\eqref{eq:like_mixture}
multiplies per-galaxy distributions that have each already been marginalised over the density
posterior samples $s$, so $\prod_i\langle\mathcal{L}_i\rangle_s\neq\langle\prod_i\mathcal{L}_i\rangle_s$
and common environmental information could be over-counted. This is unlikely to be significant in practice: the galaxies' mean separation, $14$--$16$\,Mpc, is about twice the $\lambda_C=8$\,Mpc of Eq.~\eqref{eq:headline}, so the $\sim\!2\times10^4$
independent $\lambda_C$-sized cells in the survey volume outnumber the galaxies by a factor of several. In the massless limit, where $\vec{a}_5$ is coherent across the volume, what enters the
likelihood is its projection onto each disk's own sky-plane direction perpendicular to the major axis; orientations are
effectively random, so a common $\vec{a}_5$ still yields near-independent predicted signs.
\item The reconstruction assumes $\Lambda$CDM growth, strictly speaking making the bound a consistency test. A fifth force at the level of our bound is, however, too weak to modify the growth of structure appreciably at the precision of the reconstruction, so this is a good approximation in practice.
\end{itemize}

Finally, we have taken the force on the dark matter to be the static Yukawa exchange of Eq.~\eqref{eq:yukawa} (Sec.~\ref{sec:warpmodel}). In coupled quintessence the background field also rolls, and since the dark-matter mass depends on it, the dark-matter Euler equation acquires a further acceleration $-\beta\kappa\dot\varphi\,\vec{v}_{\rm DM}$, with $\kappa=1/M_{\rm Pl}$ and $\vec{v}_{\rm DM}$ the dark-matter peculiar velocity. The baryons do not feel this term and it does not vanish in equilibrium, so it displaces the disk exactly as $\vec{a}_5$ does. In
linear theory $\vec{v}_{\rm DM}=[2f/3\Omega_m H]\,\vec{g}$ while $\vec{a}_5=(1-f_b)\vec{g}$ in the
massless limit, with $f$ the linear growth rate, $\vec{g}$ the Newtonian gravitational acceleration and $f_b$ the cosmic baryon fraction, so the two are collinear and their ratio is
$[2f/3\Omega_m(1-f_b)](\kappa\dot\varphi/H)/2\beta$. With
$\kappa\dot\varphi/H=\sqrt{3(1+w_\varphi)\Omega_\varphi}$ that is $0.10/\beta$ at $w_\varphi=-0.99$ and
$0.30/\beta$ at $w_\varphi=-0.90$: for a scalar that is genuinely rolling the neglected term
exceeds the static exchange at the couplings of interest, and is linear rather than quadratic
in $\beta$. It falls to a $21\%$ correction only as $w_\varphi\to-1$, where $\dot\varphi$ is sourced
by the coupling alone. The two could cancel only near $\beta\simeq0.2$ --- or $\Delta G/G_N\simeq0.08$, some 50 times above Eq.~\eqref{eq:headline} --- so neglecting the term can only weaken our limit. By this linear-theory
estimate, including it would tighten the coupled-quintessence bound to
$|\beta|\lesssim4\times10^{-3}$ at $w_\varphi=-0.95$.

\section{Comparison to the literature and implications}\label{sec:implications}

The tightest previous constraints on this coupling are cosmological. For a scalar lighter than $H_0$, Planck together with baryon acoustic oscillations gives $\Delta G/G_N<5.4\times10^{-3}$ \cite{Archidiacono2022}, and adding the full shape of the Baryon Oscillation Spectroscopic Survey (BOSS) galaxy power spectrum gives $4.8\times10^{-3}$ \cite{Bottaro2024} (Table~\ref{tab:bounds}(c)).
Ours is three times stronger. Previous astrophysical bounds are weaker by about two orders of magnitude: ours is 43, 61 and 360 times stronger than the Euler-equation test of \cite{Bonvin2025}, the Sagittarius stream \cite{Kesden2006,Kesden2006b} and Fornax \cite{Zamlung2026} respectively (the Sagittarius limit is additionally qualified by the fact that a strong long-range force need not render the stream asymmetric if stars and dark matter separate before disruption \cite{Keselman2009}).
The range-resolved bound of \cite{Bean2008}, $5\%$ of gravity at $\lambda_C\simeq10$\,Mpc, is 30 times weaker than ours. At the opposite extreme, the Bullet Cluster constrains dark forces of range below $\sim\!10$\,kpc \cite{Bogorad2025}, a regime disjoint from ours.

These cosmological bounds rest on a different signal from ours. The mean dark-matter density sources the scalar just as a halo does, so a scalar lighter than $H_0$ rolls throughout matter domination and the dark-matter mass changes with it. The dark-matter density then falls faster than $a^{-3}$, lowering the expansion rate by a fraction $(\Delta G/2G_N)\ln(a/a_{\rm eq})$ and altering the distances to the last-scattering surface and to the baryon acoustic oscillation tracers, which supply most of the constraining power. The full shape of the power spectrum adds little, because the enhanced growth of structure is degenerate with galaxy bias \cite{Archidiacono2022,Bottaro2024}. A heavier scalar, with $H_0<m_\varphi\lesssim H_{\rm eq}$, is displaced by the dark matter during matter domination and begins to oscillate once the Hubble rate falls below its mass. The oscillations then act as an additional matter component that does not cluster below the scalar's Jeans scale, raising the late-time matter density and suppressing the growth of structure \cite{Bottaro2025}. This gives the tightest cosmological bounds, $\Delta G/G_N\lesssim10^{-3}$ for $8\lesssim\lambda_C/{\rm Mpc}\lesssim500$ with a minimum of $7\times10^{-4}$ at $\lambda_C\simeq45$\,Mpc, rising to $2.5\times10^{-3}$ at $0.4$\,Mpc (shown over $0.4$--$8$\,Mpc in Fig.~\ref{fig:exclusion}).\footnote{\cite{Bottaro2025} give this bound only as a 95\% upper-limit curve against $m_\varphi/H_0$ (their Fig.~1), which we digitise to better than $3\%$ in $\Delta G/G_N$. The curve joins fits at about a dozen masses, of which those between $0.4$ and $44$\,Mpc are at $\lambda_C=0.44$, $4.4$ and $44$\,Mpc, so the value at $8$\,Mpc is interpolated. They caution that their modelling of nonlinear clustering is less reliable at $\lambda_C\lesssim0.4$\,Mpc.} Between $0.4$\,Mpc and ${\sim}1$\,Gpc these are tighter than our conservative bound, by a factor $1.7$ at $8$\,Mpc, and comparable to our fiducial bound above ${\sim}1$\,Mpc.

Both cosmological signals are fixed by $\Delta G/G_N$ only in the minimal model, with a quadratic scalar potential, a negligible initial displacement of the scalar (which would otherwise change the effective coupling over cosmic time \cite{Archidiacono2022}) and a coupling that has been constant since matter--radiation equality, which supplies the logarithmic enhancement. That the bound of \cite{Bottaro2025} is seven times tighter at $\lambda_C\simeq45$\,Mpc than at infinite range shows that it constrains mainly the energy density of the mediator rather than the force it exerts. Our bound requires none of these assumptions: it tests the force acting on dark matter in the present-day Universe, independently of the scalar's cosmological history. The two are therefore complementary. A coupling that grows at late times would weaken the cosmological bounds but not ours, while one that has decayed by today would do the reverse.

Within the minimal model, forthcoming galaxy surveys are forecast to reach $\Delta G/G_N\simeq2\times10^{-3}$ for a scalar lighter than $H_0$ \cite{Bottaro2024} and $4\times10^{-4}$ at ranges of tens to hundreds of Mpc \cite{Bottaro2025}, and the Euler-equation test to reach $2\%$ with Rubin and DESI \cite{Bonvin2025}. The gain from studying galaxies rather than cosmology is qualitative rather than statistical, and is not redundant with cosmology even in principle: a dark-sector fifth force and a modification of gravity are degenerate in redshift-space distortions and lensing \cite{BonvinPogosian2023}, whereas an equilibrium displacement inside a galaxy responds to
the fifth force and not to a modification of gravity that acts equally on stars and dark matter.

Several models predict a coupling in or near the range we exclude. The DESI inference of \cite{Clini2026}, which applies the model and assumptions of \cite{Bottaro2024} to DESI galaxy clustering in place of the BOSS data behind that paper's upper limit, lies $2.2$--$4.2$ times above Eq.~\eqref{eq:headline} in central value and $1.3$--$2.9$ times at the lower edge of its 68\% intervals (Table~\ref{tab:bounds}(c)). The swampland model of \cite{Agrawal2019} identifies the distance-conjecture tower with the dark sector, the dark-matter mass falling with the scalar as $m_{\rm DM}\propto\exp(-\tilde{c}\,\varphi/M_{\rm Pl})$, so that $\tilde{c}$ is our $|\beta|$ and $\Delta G/G_N=2\tilde{c}^{2}$. Its best fit, with the local $H_0$ measurement included, is $\tilde{c}\simeq0.30$, or $\Delta G/G_N=0.18$. A fit in the same framework to DESI DR2 and supernovae gives $\tilde{c}=0.05\pm0.01$ \cite{Bedroya2025}, or $\Delta G/G_N=5.0\times10^{-3}$.
We disfavour this at $2.3\sigma$, the DESI inferences of Table~\ref{tab:bounds}(c) at $1.8$ to $3.1\sigma$ and the coupled dark energy fit at $2.1\sigma$, and exclude the swampland model by two orders of magnitude (using our conservative envelope; in the fiducial model these are $\lesssim0.4\sigma$ larger for the DESI inferences, and rise to $4.6\sigma$ for the coupled dark energy fit).
The couplings these proposals prefer could therefore only evade our test by means of a screening mechanism (Sec.~\ref{sec:screening}).

\section{Conclusion}\label{sec:conclusion}

The equivalence principle is tested to parts in $10^{13}$--$10^{15}$ for baryons, while for dark matter the tightest published tests reach only parts in $10^{3}$.
Measuring the warps of 3968 edge-on disks and modelling their response to a force acting on dark matter alone,
we have shown that galaxy morphology bounds a dark-sector fifth force at $\Delta G/G_N<7.7\times10^{-4}$ (95\% credibility) for $\lambda_C\gtrsim8$\,Mpc, and $<1.7\times10^{-3}$ after inflation by the systematics of Table~\ref{tab:bounds}(b).
The bound tightens towards the longest range we can compute, so adopting these bounds for the massless limit is conservative.
For an effectively infinite range the conservative bound is a factor three below the tightest cosmological bound on the same coupling \cite{Bottaro2024} and about two orders of magnitude below astrophysical ones. At ranges between $0.4$\,Mpc and ${\sim}1$\,Gpc a cosmological analysis of the minimal model gives a tighter bound than our conservative one \cite{Bottaro2025}, but that bound rests mainly on the energy density of the oscillating scalar. Ours constrains the force acting on dark matter today, independently of the scalar's cosmological history, and is to our knowledge the tightest such constraint on interacting dark matter, and more generally on any conformally coupled dark-sector force, of range above a few Mpc.

Thin-shell screening does not rescue a cosmologically relevant coupling. Where the field tracks the minimum of its effective potential (e.g.\ Hu--Sawicki $f(R)$), screening the galaxies that set the bound would confine the force to ranges of tens of Mpc, rendering it cosmologically irrelevant. A model that escapes this relation while leaving linear density modes unscreened would weaken our bound by only a factor $1.66$.
A scalar coupled to dark matter alone with a coupling above our bound therefore requires Vainshtein screening or a disformal coupling. Fits of interacting dark energy should impose $\Delta G/G_N<1.7\times10^{-3}$ as a conservative prior on the conformal coupling for consistency with galaxy morphology, unless the model evades the test by one of these mechanisms.

\section*{Data and code availability}\label{sec:data_avail}

The code is available at \url{https://github.com/harrydesmond/dark-fifth-force-warps} and the data products at \href{https://doi.org/10.5281/zenodo.23343105}{doi:10.5281/zenodo.23343105}.

\begin{acknowledgments}
We thank Carlos Garc\'{\i}a-Garc\'{\i}a, Alex Roskill and William Wolf for interesting discussions. HD is supported by a Royal Society University Research Fellowship (grant no.~211046). PGF
acknowledges support from STFC and the Beecroft Trust. 
\end{acknowledgments}

\appendix

\section{Extrapolation to the massless limit}\label{app:massless}

Here we justify in more detail the extrapolation of the $8$\,Mpc bound to the (quasi-)massless $\lambda_C\gtrsim4$\,Gpc regime. The isotropic variance of the
fifth-force field is
\begin{eqnarray}\label{eq:kernel}
\left\langle|\vec{a}_5|^{2}\right\rangle\propto
    \int\!\frac{d^3k}{(2\pi)^3}\,P_{\rm DM}(k)\,\frac{k^{2}}{(k^{2}+m_\varphi^{2})^{2}},
\end{eqnarray}
    and
\begin{eqnarray}\label{eq:kernel2}  
  \frac{\partial}{\partial m_\varphi^{2}}\frac{k^{2}}{(k^{2}+m_\varphi^{2})^{2}}<0
    \quad\text{for all }k ,
\end{eqnarray}
so the kernel, and with it the mean-squared field, rises at every wavenumber as the mass falls. That is however an isotropic ensemble average, whereas the warp responds to the projection of $\vec{a}_5$ onto the disk normal, in which distant structure could cancel. Thus a larger predicted amplitude does not by itself tighten a limit, which depends also on the predicted signs and their chance correlation with the measured warps.

We therefore recomputed the forward model over the whole sample at $\lambda_C=8$, $12$, $16$ and $20$\,Mpc, five times at each range, and evaluated Eq.~\eqref{eq:like_mixture} against the measured warps of the 3968 galaxies of Sec.~\ref{sec:data}.\footnote{For computational expediency we use 300 Monte Carlo realisations and two density samples rather than 1000 and 10 as in \DF{}, and freeze the regressors for the halos of faint galaxies (Sec.~\ref{sec:inputs}) at their $\lambda_C=8$\,Mpc training. The five $8$\,Mpc limits in this configuration average $0.93$ of the fiducial one. Since $\lambda_C=8$\,Mpc is rerun identically, the ratios are internally consistent, and the predicted signals reproduce those of \DF{} to $1.5$--$3\%$ at every percentile.} Each run yields its own $95\%$ upper limit on $\Delta G/G_N$. These scatter by $11\%$ between the runs at $8$\,Mpc and by $25\%$ at $20$\,Mpc, so we compare runs in pairs: four of the runs at each longer range share their random seeds with four of those at $8$\,Mpc, so that the Monte Carlo noise common to a pair cancels in its ratio. The ratio of the limit to its $8$\,Mpc partner is $0.90\pm0.10$, $0.83\pm0.09$ and $0.76\pm0.05$ at $12$, $16$ and $20$\,Mpc (mean and standard deviation over the four pairs), below unity at $p=0.07$, $0.02$ and $0.002$ on one-sided paired $t$ tests. The longer-range data therefore exclude smaller couplings, and the bound strengthens towards larger range, continuing the trend over $0.4$--$8$\,Mpc shown in Fig.~\ref{fig:exclusion}.
Adopting the $8$\,Mpc limit at every larger range, as we do in the main text, is therefore conservative.

Beyond $10$\,Mpc the source sphere of radius $4\lambda_C$ within which we include fifth-force contributions extends past the edge of the reconstruction for the most distant galaxies, where a one-sided sum would bias $\vec{a}_5$ radially towards the observer and could itself fake the tightening just reported. Splitting the sample by distance rules that out: if truncation were responsible, the galaxies with the least complete source spheres would show the effect most strongly, but they show it least.
The 1409 galaxies beyond $200$\,Mpc give paired ratios of $0.95$, $0.86$ and $0.81$, while the 1334 within $150$\,Mpc, whose source spheres are complete to $25$\,Mpc, give $0.84$, $0.71$ and $0.62$. The tightening is therefore not an artefact of the reconstruction's finite extent:
it is strongest exactly where the calculation is most trustworthy.

\section{Translation between conventions}\label{app:conv}

Three conventions for the fifth-force coupling are in use in the literature, related by
\begin{equation}\label{eq:conventions}
  \frac{\Delta G}{G_N}=2\beta^{2}
  =\frac{\kappa_Y^{2}}{4\pi G_N m_{\rm DM}^{4}}
  =\frac{G_s}{4\pi G_N}\equiv\beta_{\rm Clini}.
\end{equation}
Here $\beta$ is the coupling used in the main text, normalised as in \cite{CoupledDESI2026} (\cite{Amendola2000} defines $\beta$ such that $\Delta G/G_N=4\beta^2/3$),
$\kappa_Y\equiv m_{\rm DM}\,\partial m_{\rm DM}/\partial\varphi$ the Yukawa charge of a light mediator
\cite{BuckleyFox2010} and $G_s$ the scalar-exchange coupling of
\cite{Clini2026}.
$\beta$ and $\beta_{\rm Clini}$
are related by $\beta=\sqrt{\beta_{\rm Clini}/2}$, so that $\beta_{\rm Clini}=0.0037$ is
$\beta=0.043$, a factor $1.4$ above the $|\beta|\simeq0.03$ of contemporaneous coupled-quintessence fits
\cite{CoupledDESI2026}, while at our own bound the two differ by a factor $17$.

To derive Eq.~\eqref{eq:conventions}, start from the underlying action
\begin{equation}\label{eq:action}
\begin{aligned}
  S&=\int\! d^4x\sqrt{-g}\left[\frac{R}{16\pi G_N}-\tfrac12(\partial\varphi)^2-V(\varphi)\right]\\
   &\quad+S_{\rm DM}\!\left[\psi_{\rm DM};g_{\mu\nu},\varphi\right]+S_b\!\left[\psi_b;g_{\mu\nu}\right],\\
  m_{\rm DM}(\varphi)&=\bar{m}_{\rm DM}\,e^{\beta\kappa\varphi},\qquad \kappa^2=\frac{1}{M_{\rm Pl}^2}=8\pi G_N ,
\end{aligned}
\end{equation}
with $R$ the Ricci scalar, $V$ the scalar potential, $S_{\rm DM}$ and $S_b$ the dark-matter and
baryonic actions, and $m_\varphi^2=V''$ at the background field value. Varying with respect to
$\varphi$, with the dark matter a pressureless gas of mass density $\rho_{\rm DM}$ and
$\partial m_{\rm DM}/\partial\varphi=\beta\kappa\,m_{\rm DM}$, and linearising about the
background field, gives $(\nabla^{2}-m_\varphi^{2})\delta\varphi=\beta\kappa\,\rho_{\rm DM}$. The
Green's function $-e^{-r/\lambda_C}/4\pi r$ then puts a point source of mass $M$ at
$\delta\varphi=-\beta\kappa M e^{-r/\lambda_C}/4\pi r$, and a second particle in that field has rest
energy $m_{\rm DM}(1+\beta\kappa\,\delta\varphi)$, hence interaction energy
\begin{equation}\label{eq:apppot}
\begin{aligned}
  V_5(r)&=\beta\kappa\,m_{\rm DM}\,\delta\varphi(r)
        =-\frac{\beta^{2}\kappa^{2}}{4\pi}\frac{m_{\rm DM} M}{r}e^{-r/\lambda_C}\\
        &=-2\beta^{2}\,\frac{G_N m_{\rm DM} M}{r}\,e^{-r/\lambda_C} ,
\end{aligned}
\end{equation}
the last step using $\kappa^2=8\pi G_N$. Comparison with $-G_N m_{\rm DM} M/r$ gives
$\Delta G/G_N=2\beta^2$: the factor of two is one power of $\beta\kappa$ at the source and
one at the test particle, while the $4\pi$ of the Green's function cancels against the $8\pi$ of
$\kappa^2$, so no stray $4\pi$ survives. Only the first derivative of $m_{\rm DM}$ enters, so a general
$\varphi$-dependent mass gives
$\Delta G/G_N=(\partial m_{\rm DM}/\partial\varphi)^2/(4\pi G_N m_{\rm DM}^{2})$, the second equality of
Eq.~\eqref{eq:conventions}. Since $G_s$ is defined in \cite{Clini2026} through the potential
$-G_s m_{\rm DM}^2/4\pi r$, $\beta_{\rm Clini}=G_s/4\pi G_N$ is $\Delta G/G_N$ itself.

\section{Likelihood evaluation and mock tests}\label{app:neglected}

We evaluate the posterior of Eq.~\eqref{eq:like_mixture} on a fixed grid in $\Delta G/G_N$, with spacing $2\times10^{-6}$ below $6\times10^{-4}$, $10^{-5}$ to $3\times10^{-3}$, $10^{-4}$ to $2\times10^{-2}$ and $5\times10^{-4}$ to $0.15$. At each grid point we maximise the likelihood over $\sigma_{w_1}$ and $\ln\nu$, restarting from several initial values, and integrate over them in the Laplace approximation. Quantiles are taken from a cumulative trapezoid integral: a plain cumulative sum gives the $\Delta G/G_N=0$ endpoint full rather than half weight which, since the posterior peaks at that boundary, biases upper limits low. No grid point with appreciable posterior mass has a failed maximisation, and less than $10^{-4}$ of the mass lies in the top 5\% of the grid in every analysis we report. The alternative noise model of Table~\ref{tab:bounds}(b), a mixture of a Gaussian and a Student-$t$ distribution, which contains the fiducial model as a special case, is preferred by $\Delta\ln Z=5.1$ and gives a bound tighter by a factor $0.88$, so the fiducial model is the conservative choice.

We test the inference on mock data at $\lambda_C=8$\,Mpc, with 20 mocks at each of $\Delta G/G_N=0$, $10^{-4}$, $2\times10^{-4}$, $4\times10^{-4}$ and $10^{-3}$. Each galaxy's true warp is drawn from its predicted-signal distribution and its noise from one of four models: the fitted Student-$t$, a Gaussian core with Gaussian outliers that replace the signal, and a Gaussian core with outliers resampled from the measured warps beyond 3.5 times their scatter that either replace the signal or are added to it. The 95\% upper limit covers the injected value in 95--100\% of mocks at every level and for every noise model. At $10^{-3}$ the median of the posterior medians is $(1.02$--$1.26)\times10^{-3}$, and at lower levels it lies above the injected value, as expected for a posterior bounded at $\Delta G/G_N=0$. With no injected signal the median 95\% limit is $8.2\times10^{-4}$ to $1.1\times10^{-3}$. The upper limits are therefore not biased low.

Mock data with no fifth force also calibrate the excisions of Sec.~\ref{sec:systests}. At each of the twenty force ranges we generate 40 such data sets, 20 with warps drawn from the fitted Student-$t$ distribution and 20 from a Gaussian core with outliers resampled from the measured warps, apply both excisions to each and record the ratio of the excised to the full-sample 95\% limit. The median of this ratio over the 40 mocks is the weakening expected from the loss of the most informative galaxies alone---an excision enters the conservative bound only through the factor by which its ratio on the real data exceeds this.

\bibliographystyle{apsrev4-2}
\bibliography{refs}

\end{document}